\documentclass[conference]{IEEEtran}

\usepackage[letterpaper, left=1in, right=1in, bottom=1in, top=0.75in]{geometry}
\usepackage{graphicx}
\usepackage{amsmath}
\usepackage{amsfonts}
\usepackage{subfigure}
\usepackage{multirow}
\usepackage{stfloats}
\usepackage{cite}
\usepackage{color}
\usepackage{graphicx,subfigure,float}
\usepackage{balance}

\usepackage{epsf}

\usepackage{hyperref}

\begin{document}


%
\title{Grant-less Uplink Transmission for LTE Operated in Unlicensed Spectrum}

\author{\IEEEauthorblockN{Jinyu Zhang$^1$, Wenting Chang$^2$, Huaning Niu$^2$, {Salvatore Talarico$^2$}, Hongwen Yang$^1$}
\IEEEauthorblockA{$^1$Beijing University of Posts and Telecommunications\\
$^2$Intel Mobile Communications Technology Ltd.\\}
Email: zhangjinyu@bupt.edu.cn
}

\maketitle

\begin{abstract}
Deployment of Long Term Evolution (LTE) in unlicensed spectrum has been a candidate feature  to meet the explosive growth of traffic demand since 3GPP release 13. To further explore the advantage of unlicensed bands, in this context the operation of both uplink and downlink has been supported and studied in the subsequent releases. However, it has been identified that scheduled uplink transmission performance in unlicensed spectrum is significantly degraded due to the double listen-before-talk (LBT) requirements at both eNB when sending the uplink grant, and at the scheduled UEs before transmission. In this paper, in order to overcome this issue, a novel uplink transmission scheme, which does not require any grant, is proposed, and the details regarding the system design are provided. By modeling the dynamics in time of the LBT for both a system that employs a conventional uplink scheme, as well as the proposed scheme, it is verified through analytical evaluation that the double LBT scheme for uplink transmission greatly reduces the channel access probability for the UE, and leads consequently to performance loss, while the proposed scheme is able to alleviate this issue. System level simulation results, compliant with the LTE standard, show that the proposed scheme can achieve a significant performance gain in terms of throughput with negligible performance loss for the downlink, and other technologies operating in the same spectrum.
\end{abstract}

\IEEEpeerreviewmaketitle

\section{Introduction}
Owing the popularity of smartphones, tablets, and other wireless devices, the recent widespread adoption of wireless broadband has resulted in a tremendous growth in the volume of mobile data traffic, which is projected to continue unabated \cite{bio1_Cisco,bio2_Nokia_WP}. As a consequence of this, the system capacity of wireless communication systems have been severely challenged. However, restricted by the lack of available spectrum resource in licensed band, the traditional Long Term Evolution (LTE) technology is powerless in tackling this problem. Therefore, the available resources in unlicensed band have attracted recently more, and more attention as an important complement to alleviate the high data traffic load \cite{bio3_HW_WP,bio4_Q_WP}. In this regards, 3GPP has introduced its operation in unlicensed band via Licensed Assist Access (LAA) in release 13 \cite{bio5_LAA,bio6_TR889}. LAA uses carrier aggregation in the downlink to combine LTE in unlicensed spectrum with LTE in the licensed band to expand the system bandwidth. {While significant changes have been made compared to the LTE framework through the introduction of several mechanisms \cite{bio_new2_LTEU_LAA}, the authors of \cite{bio_new1_LAA_Rel13} have shown that LAA ensures fair coexistence with existing Wi-Fi networks.}


{Following the current momentum on unlicensed spectrum, recently 3GPP has started two new working items, named ``new radio (NR) based unlicensed access" \cite{bio_new3_NRSI} and ``Enhancements to LTE operation in unlicensed spectrum" \cite{bio_new4_NRWI}.} With this in mind, apart from LAA systems,  MulteFire (MF) systems \cite{bio7_MF} employ LTE technology, but solely work in unlicensed spectrum without assistance of the ``anchor" in licensed spectrum. For these systems  the control information and reference signal must be supported to be transmitted on unlicensed carriers along with the entire data. In this regards, a MF system is totally different than an LAA system, and its system framework needs to be modified to support the sole operation in unlicensed band. Even though the MF technology is still at an embryonic stage, the combination of LTE like performance benefits, and WiFi like deployment simplicity makes MF a significantly important supplement, and valuable study topic to meet the ever-increasing wireless traffic.

\begin{figure*}[!t]
\centering
\includegraphics[width=6.3in, height=2.0 in]{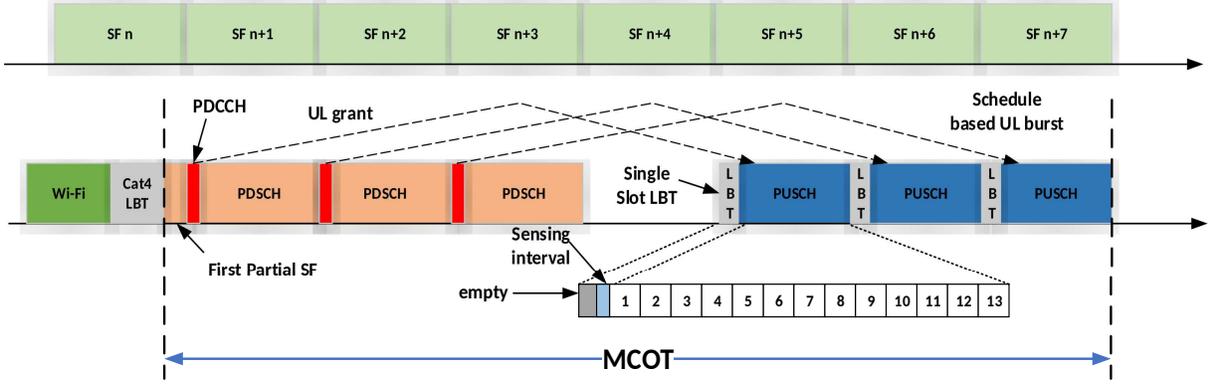}
\vspace{-0.35cm}
\caption{{Scheduled based uplink transmission mode}}
\label{SUL_mode}
\vspace{-0.25cm}
\end{figure*}

\newcounter{mytempeqncnt}
\begin{figure*}[hb]
\normalsize
\setcounter{mytempeqncnt}{\value{equation}}
\setcounter{equation}{0}
\hrulefill
\begin{equation}\label{eq:1}
\begin{split}
p_{tx}^{WiFi}&=\frac{2q(1\!-\!p_b)(1\!-\!2p_f)}{2(1-p_b)(1-p_f)(1-2p_f)+q[W_0p_f(1-(2p_f)^m)+(1+W_0-2p_b)(1-2p_f)]}\\
\end{split}
\end{equation}
\begin{equation}\label{eq:1b}
\begin{split}
p_{tx}^{Cat4}&=\frac{2q(1\!-\!p_b)(1\!-\!p_f)R}{Q\!+\!q[W_0P(1\!-\!p_f)(1\!-\!(2p_f)^{(m\!+\!1)})\!+\!PR(1\!-\!2p_b)(1\!-\!2p_f)\!+\!2R(1\!-\!p_b)^2(1\!-\!p_f)(1\!-\!2p_f)]}
\end{split}
\end{equation}
\setcounter{equation}{\value{mytempeqncnt}}
\vspace*{1pt}
\end{figure*}
\setcounter{equation}{2}

In legacy LTE systems, scheduled based uplink (SUL) transmission has been considered in the 3GPP release 14 study, wherein uplink transmission is conditional to an explicit uplink grant via physical downlink control channel (PDCCH) \cite{eLAA}. In order to comply with the FCC regulation requirements, and in order to maintain fair coexistence with other technologies, the listen-before-talk (LBT) mechanism is applied to check whether the channel is clear or occupied before using it. However, the use of the legacy two stage modality for LBT reduces the uplink channel access probability. This drawback is highlighted and verified in terms of channel access probability by modeling the dynamics in time of a system employing LBT throughout a Markov chain, similarly as \cite{bio8_Bianchi,bio9_CChen,bio10_MarkovChain}. Besides the penalty imputable to a lowered channel access probability, the performance of SUL is also negatively affect by the processing delay (generally 4 ms due to hardware constraints) between the uplink grant and the scheduled transmission, which may also lead to transmission latency and resource waste in case there is no downlink data. Hence, scheduled based uplink transmissions are not suitable in unlicensed band.

In this paper, we consider a new uplink transmission scheme, which does not require any eNB scheduling grant, named \textit{grant-less uplink (GUL)} transmission. While this methodology highly resembles that currently used in the WiFi uplink design, it is a significant departure from the existing SUL transmission of the legacy LTE. In this regards, a number of enhancements, which are discussed along this manuscript, need to be made with respect to the legacy LTE design in order to be able to properly enable and perform GUL transmissions.

The rest of this paper is organized as follows. Section \ref{section_2} begins with a brief introduction of the SUL scheme. This section continues by building an analytical framework based on a Markov chain model, which is employed to model the dynamics in time of the LBT procedure for both the user equipment (UE) and eNB. This analytical framework is then used to compare the channel access probability of SUL and GUL schemes, and highlight the benefits of the proposed scheme. The overview procedure and design details for GUL mode are then provided in section \ref{section_3}. In section \ref{section_4}, the performance of the proposed scheme is evaluated via system level simulations. Finally, conclusions are drawn in section \ref{section_5}.

\section{System Model}   \label{section_2}

In legacy LTE, the UE that intends to transmit data needs to obtain an uplink grant from the serving eNB, and only then it can start uplink transmission, as illustrated in Fig.\ref{SUL_mode}. In primis, the eNB is required to perform Cat.4 LBT on the target carrier for the uplink grant transmission, as regulated in \cite{bio11_ETSI893,bio12_TS213}. Once it is able to access successfully the channel, the subsequent maximum channel occupancy time (MCOT) can be occupied, and scheduled for either downlink or uplink transmission by the eNB. While the PDCCH carrying uplink grant can be transmitted in the first available subframe (SF), due to the processing delay the physical uplink shared channel (PUSCH) is scheduled at the latter SFs during the same MCOT. The remaining symbols in the downlink SFs can be utilized for downlink data transmission, if any. Before uplink transmissions can take place, the scheduled UE needs to complete an additional LBT (either single interval LBT or Cat.4 LBT) \cite{bio12_TS213} after receiving the grant. If this second LBT fails, the resources reserved for uplink are wasted.

Intuitively, the SUL mode hampers the channel access probability for the UE. In order to overcome this issue, it is proposed here to adopt one-LBT uplink access mechanism instead of this double-LBT procedure, which leads to uplink transmissions that can be performed autonomously without requiring grants, which we refer to as GUL. For the proposed GUL scheme, on the other hand, similarly to SUL, Cat.4 LBT is still employed for the fair sharing of unlicensed band.

{While Markov chains and their properties have been extensively used to model and characterize the procedure of LBT for Wi-Fi and LAA \cite{bio8_Bianchi,bio9_CChen,bio10_MarkovChain,bio_new5_ThvLAA}, in this contribution they are used to model the LBT for MF systems, in order to study its coexistence with Wi-Fi. In particular, utilizing} a Markov chain model, the LBT procedure of a WiFi, and MF access node is modeled, and the transmit probabilities  in a randomly chosen slot time can be calculated by (\ref{eq:1}) and (\ref{eq:1b}), respectively \cite{bio9_CChen,bio10_MarkovChain}. For these equations,  $Q=2(1-p_b)(1-p_f)(1-2p_f)$, $P=(p_b+p_f-p_bp_f)$, $R=(1-p_f^{(m+1)})$, $q$ denotes the probability of packet arrival, $m$ is the maximum clear channel assessment (CCA) stage, and $W_0$ is the initial contention window size. $p_f$ and $p_b$ denote the probability of transmission failure due to collisions, and the probability that the channel is detected to be occupied, respectively. In \cite{bio9_CChen,bio10_MarkovChain}, these probabilities are although determined under the simplified assumption that all the nodes in the coexistence scenario can detect the signal from all other nodes above the carrier sense threshold. In order to address this issue, the distribution of the detected energy \cite{bio13_ED1,bio14_ED2,bio15_LBT} is here taken into account. For simplicity, let assume that the path loss between any two nodes are identical. The total receiving power $P_{rx}$ can be then be obtained by multiplying the receiving power $P_{0rx}$ from a single transmitting node by the total number of the transmitters $n$. Thus, the distribution of the energy detection conditioned on the number of transmitters could be expressed as follows
\begin{equation}\label{eq:2}
\hspace{-0.1 cm}f_Y(y|n)=
\begin{cases}
\frac{1}{2^\mu\varGamma(\mu)}y^{\mu-1}\mathrm{e}^{-\frac{y}{2}}&   \text{idle}\\
\frac{1}{2}(\frac{y}{2\gamma})^{\frac{\mu-1}{2}}\mathrm{e}^{-\frac{2\gamma+y}{2}}I_{\mu-1}(\sqrt{2\gamma y})&   \text{busy}\\
\end{cases}
\end{equation}
where $\gamma$ is the signal-to-noise ratio (SNR), and depends on the number of transmitters $n$ since $\gamma= nP_{0rx}/P_{noise}$, $\varGamma(.)$ represents the gamma function, and $I_v(.)$ is the $v$th-order modified Bessel function of the first kind.

Assume that the channel sense failure and the transmission collision both occur in the case that the detected energy is above the LBT threshold $y_{thv}$. In a network with $N$ access nodes, it yields that
\begin{equation}\label{eq:3}
p_b\hspace{-0.08cm}=\hspace{-0.08cm}p_f\hspace{-0.08cm}=\hspace{-0.1cm}\sum_{n=1}^{N-1}\hspace{-0.1cm} \dbinom{N}{n} p_{tx}^n(1\hspace{-0.08cm}-\hspace{-0.08cm}p_{tx})^{N\hspace{-0.05cm}-\hspace{-0.05cm}1\hspace{-0.05cm}-\hspace{-0.05cm}n} \hspace{-0.15cm}\int_{y_{thv}}^{+\infty}\hspace{-0.25cm} f_Y(y|n)\,dy.
\end{equation}

The transmission probability for a WiFi Access Point (AP), and a system with Cat.4 LBT is evaluated then by solving (\ref{eq:1}) and (\ref{eq:1b}) using (\ref{eq:3}), respectively. For a SUL scheme, only $N$ eNBs perform Cat.4 LBT, and the uplink channel is available only when both downlink Cat.4 LBT, and the single slot LBT at the UE side succeed. Thus, in this case the uplink channel access probability can be expressed as
\begin{equation}\label{eq:4}
p_{tx}^{SUL}=(1-p_b)p_{tx}^{Cat4}
\end{equation}

In the proposed GUL mode, the UE performs independent Cat.4 LBT, which is nearly the same behaviour as eNB in respect to the channel access procedure. Therefore, the channel access probability for both UE and eNB can be obtained by substituting within $N$ in (\ref{eq:3}) the sum of the number of UEs and eNBs involved.

\begin{figure}[!t]
\centering
\includegraphics[width=7.00cm]{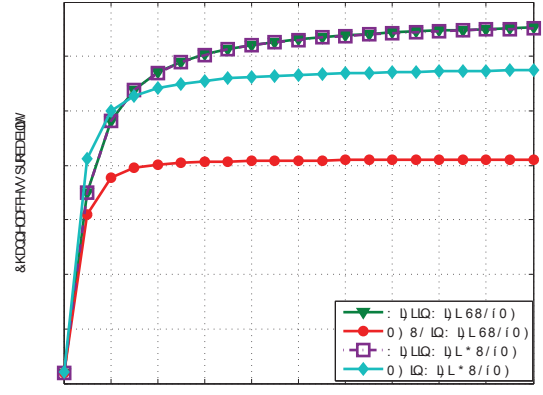}
\vspace{-0.5cm}
\caption{Channel access probability}
\label{ChAccessPro}
\vspace{-0.25cm}
\end{figure}

\begin{figure*}[!t]
\centering
\includegraphics[width=6.3in, height=2.0 in]{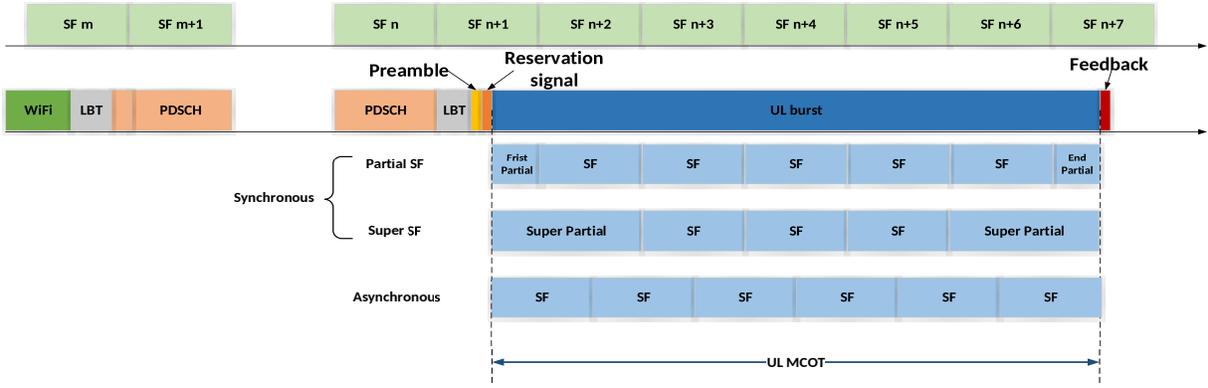}
\vspace{-0.1cm}
\caption{Grant-less uplink transmission mode}
\label{GUL_mode}
\vspace{-0.25cm}
\end{figure*}

Fig. \ref{ChAccessPro} shows the channel access probability based on the assumptions that $y_{thv}=-72$ dBm for both the Wi-Fi and the MF system. The number of WiFi APs or MF eNBs deployed for each operator is $N=5$, and each eNB only has one active UE. For this plot, $m=4$, $W_0 = 16$, $P_{0rx}/P_{noise}\approx 10$ and the following two scenarios are shown:
\begin{itemize}
\item WiFi+SUL-MF: WiFi APs of operator 1 coexist with MF eNBs and UEs of operator 2, which operates in scheduled based uplink modality;
\item WiFi+GUL-MF: WiFi APs of operator 1 coexist with MF eNBs and UEs of operator 2, which operates in grant-less based uplink modality.
\end{itemize}

 As illustrated by Fig. \ref{ChAccessPro}, the SUL scheme is subject to a small uplink channel access probability due to the double LBT required, while for the GUL mode this improves significantly with negligible impact on the WiFi system performance.

\section{System Design of Grant-less Uplink Mode} \label{section_3}

Fig. \ref{GUL_mode} provides an illustration of the overall procedure for the GUL mode. Firstly, the UE with uplink data performs channel sensing. In this case, Cat.4 LBT is adopted to maintain the fair coexistence with the incumbent system and other technologies. A preamble signal is needed before data transmission for the detection at the anchored eNB, and signalling of control information. A reservation signal may also be needed to align with the predefined boundary. Then, the UE can use the whole MCOT for data transmission rather than shared with downlink. Finally, the eNB needs to feedback the ACK/NACK information for Hybrid Automatic Repeat Request (HARQ) process.

The proper use of GUL mode needs a quite different framework than that used by SUL in the LTE technology. Therefore, a number of enhancements, such as the control information and feedback, are needed with the respect to the legacy LTE design, and the details are discussed in this section.

\subsection{Detection of PUSCH at eNB side}

Due to the lack of scheduling, the serving eNB is not aware of the UE's transmission and it needs to detect the presence of the uplink burst. Two candidate methods can be taken into consideration for such indication:
\begin{itemize}
\item Implicit indication by demodulation reference signal (DMRS): the serving eNB performs blind detection of the DMRS sequence to infer the presence of PUSCH;
\item Explicit indication through Uplink Control Indicator (UCI): in this context, the existing UCI formats can be reused to provide additional information regarding the uplink burst. The content of UCI includes but is not limit to the following fields: HARQ process number, UE identifier, and new data indicator (NDI).
\end{itemize}

\subsection{Uplink Sub-frame Design}

Since the LBT could be completed at any time instant, mostly not aligned with the primary cell (PCell) SF boundary, this may result in a waste of resources due to the fact that  the transmission is postponed until the boundary of next SF. In order to better utilize the interval of time from the ends of the LBT until the PCell sub-frame boundary, a more flexible design of the uplink SF is required. As shown in Fig. \ref{GUL_mode}, three uplink SF types can be adopted:
\begin{itemize}
\item Synchronous SF, which is aligned with the boundary of PCell SF to minimize the implementation impact. In this context, a partial SF or super SF can be defined on a subset of OFDM symbols within the uplink SF (similar to the partial TTI for downlink LAA), while  the PCell still remains aligned in terms of timing relationship with the uplink burst transmission. In this case, the UE can start PUSCH transmission at certain known OFDM symbol positions within a SF  with the aim to limit the UE scheduling complexity. In particular, as UE may know the duration of the partial TTI, it may need to create multiple potential partial SFs corresponding to different hypothesis of possible partial SF. However, this incurs in a significant computation and buffer complexity at the UE side. Thus it is desirable to limit the set of possible starting positions to assume some predefined and restrict values, e.g. \{1, 8\}.
\item Asynchronous SF, which cannot be aligned with PCell boundary as illustrated in Fig. \ref{GUL_mode}. As long as the channel is acquired through LBT, UE could carry out the uplink transmission based on the legacy 1 ms SF design.
\end{itemize}

\subsection{Scheduling, Link Adaptation and HARQ Operation} \label{LA_HARQ}

 Instead of relying on the indication from the serving eNB, the UE needs to autonomously select the resource allocation in GUL mode. Accurate channel state information (CSI) is essential for both the scheduling at the UE side, and the demodulation at the eNB side. Apart from this, the UE also needs feedback information for HARQ retransmission. In this regards, the process could be summarized in following steps:
\begin{itemize}
\item Step 1: eNB estimates and calculates the uplink CSI based on the sounding reference signals (SRSs) from the UE. In particular, in this case the legacy LTE design for SRS can be reused, and they can be transmitted in the last OFDM symbol. Additionally, the CSI request can be transmitted along with SRSs.
\item Step 2: The UE chooses an appropriate modulation and coding scheme (MCS). The selection can be done by the eNB, which can indicate the best suitable MCS to UE. Alternatively, the UE can request CSI, and based upon this information it can select the appropriate MCS by itself.
\item Step 3: The UE transmits data along with the scheduling information via PUCCH, which may contain the HARQ process number and NDI.
\item Step 4: The eNB transmits ACK/NACK feedback via PDCCH, after receiving uplink data.
\end{itemize}
For link adaptation, the possible suitable options are:
\begin{itemize}
\item The eNB dynamically feedbacks the uplink CSI for MCS selection, while indicating HARQ ACK/NACK feedback;
\item The UE uses the MCS indicated in latest DCI.
\end{itemize}

\subsection{Control Channel Design}
\begin{figure}
\centering
\includegraphics[width=7.75cm]{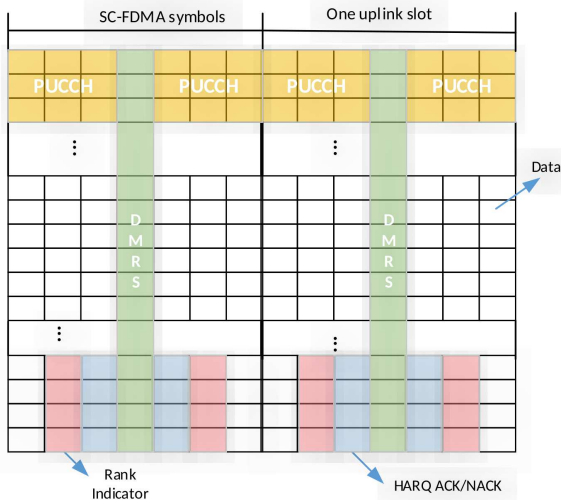}
\vspace{-0.6cm}
\caption{Illustration of PUCCH control region}
\label{PUCCH}
\vspace{-0.50cm}
\end{figure}

When UEs have simultaneous uplink data and control transmission, control signaling can be multiplexed with data prior to the discrete fourier transformation (DFT) to preserve the single-carrier property in the uplink transmission as shown in Fig.\ref{PUCCH}. This methodology can be reused in systems such as MF, which work solely in unlicensed spectrum, but with certain extent. In fact, in these systems a different content can be carried, and a list of possible fields is as follows:
\begin{itemize}
    \item Cell radio network temporary identifier (C-RNTI);
    \item HARQ process number;
    \item NDI, which is used to state whether the current transmission is a retransmission or not;
    \item MCOT and uplink burst related information described as the number of SFs. In case not all the SFs are used, the remaining SFs could be scheduled by eNB for downlink or uplink transmission of other UEs.
    \item Carrier used;
    \item A-CSI, and HARQ ACK/NACK bitmaps.
\end{itemize}

In order to reduce the UCI signalling overhead, it is preferred to transmit some of this information once, especially for fields such as A-CSI and HARQ ACK/NACK bitmaps, while some other fields which are essential (i.e, HARQ process number, C-RNTI and NDI) should be transmitted in each SF. In this regard, at least two different sizes for UCI should be predefined: one which includes the complete UCI with all fields, and one which incorporates only necessary fields. Since both the MCS index (which can be determined according to subsection \ref{LA_HARQ}) and the UCI size are needed to separate the control information, the eNB can perform blind detection to determine this last information. This option is although very computational intense, and alternatively the HARQ ACK/NACK or the rank indicator (RI) can be used to indicate the UCI size.

\begin{figure}
\centering
\includegraphics[width=7.75cm]{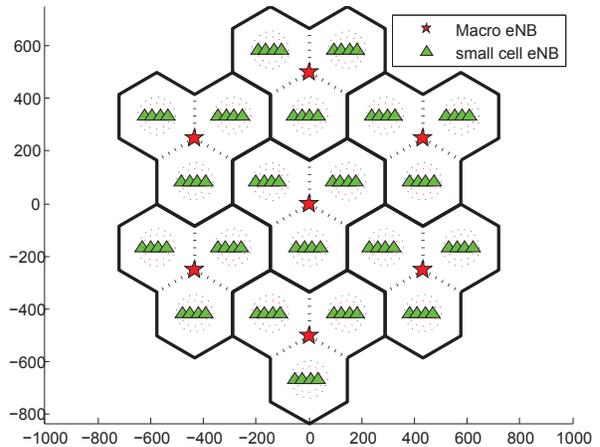}
\caption{Network layout for the indoor scenario}
\label{deploy}
\end{figure}

\begin{table}
\centering
\caption{Main Simulation Parameters}
\begin{tabular*}{2.4in}{c|c}
\hline
\hline
\textbf{Parameters} & \textbf{Value}\\ \hline
Scenario Layout & Indoor Scenario   \\  \hline
Number of UE & 20 UEs per sector \\    \hline
Channel Model & WINNER B+   \\    \hline
Carrier frequency & 5GHz\\ \hline
Inter-Station Distance &500m\\ \hline
MCOT & 5ms \\ \hline
Traffic Model & FTP Model 3  \\  \hline
File Size & 0.5MByte  \\  \hline
DL:UL Traffic Ratio & 50:50  \\   \hline
eNB Tx Power & 18dBm  \\    \hline
eNB Antenna Gain  &  5dB   \\   \hline
UE Tx Power  & 18dBm   \\   \hline
UE Antenna Gain & 0dB  \\   \hline
\end{tabular*}
\label{ParaList}
\vspace{-0.5cm}
\end{table}

\section{Simulation and Performance Evaluation} \label{section_4}

\begin{figure}
\centering
\hfill
\subfigure[Average uplink UPT]
{\includegraphics [width=7.25cm, height=1.6 in] {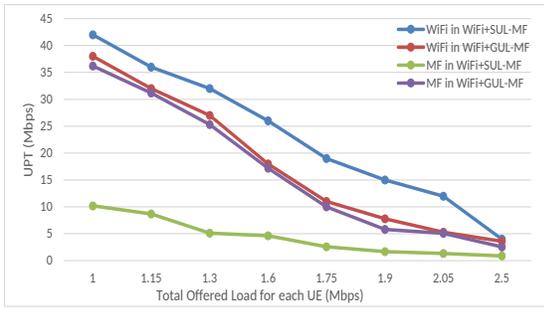} \hfill
\label{UL_UPT}
}
\hfill
\vspace{-0.10cm}
\subfigure[Average downlink UPT]
{\includegraphics [width=7.25cm, height=1.6 in] {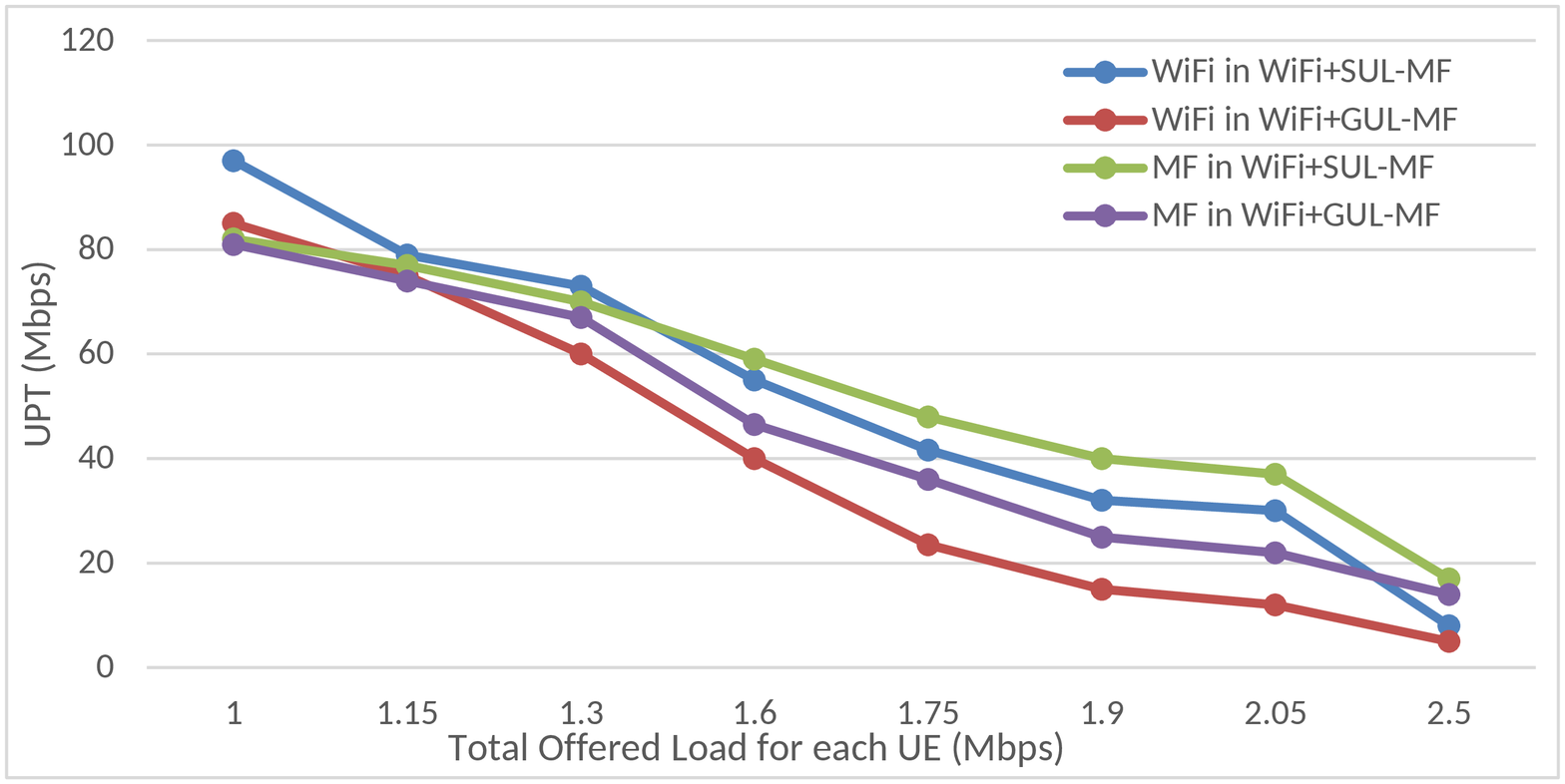}
\label{DL_UPT}
}
\caption{Average UPT performance}
\label{UPT_all}
\vspace{-0.25cm}
\end{figure}

This section provides the results obtained from comprehensive system level simulations performed with the aim to evaluate the performance of the proposed scheme. The simulations are performed under the assumptions agreed in \cite{bio6_TR889}, which are summarized in Table \ref{ParaList}. An indoor deployment with 7 hexagonal cell sites is considered for each operator, as shown in Fig.\ref{deploy} . Every site has 3 sectors with 4 MF eNBs or 4 Wi-Fi APs randomly dropped, and grouped as a cluster. Similarly to Fig. \ref{ChAccessPro} two scenarios are considered: WiFi+SUL-MF, and WiFi+GUL-MF. Fig. \ref{UPT_all} shows the performance in terms of the mean user perceived throughput (UPT) for these two scenarios for both a WiFi and MF systems and for both uplink and downlink.

This figure highlights that the uplink throughput of a MF operator is quite low when the SUL scheme is adopted for the aforementioned issues, and that the proposed GUL scheme allows to improve significantly its performance up to achieving comparable performances with WiFi. As for the downlink, the proposed scheme leads to a slight performance loss, which is negligible compared to the gain obtained for uplink. In fact, performance in terms of sum throughput of both downlink and uplink are still significantly improved with the proposed scheme. Moreover, by comparing the performances of WiFi in terms of throughput between the case when the SUL and the GUL scheme is used, it is possible to notice that the proposed scheme is able to still guarantee coexistence between MF and WiFi. For both downlink and uplink, the WiFi performance is slightly degraded due to the intense channel access competition with MF. However, such performance degradation is acceptable, and it may also incur in case the number of WiFi APs are increased in a certain area. In conclusion, the proposed GUL scheme can achieve remarkable performance gain for MF uplink and maintain the friendly coexistence with incumbent MF downlink, and Wi-Fi technology.

\section{Conclusion}  \label{section_5}
In this paper, in order to cope with the tremendous deterioration of the uplink performance of LTE systems operating in unlicensed spectrum, such as MF, a new transmission scheme is proposed, which allows to perform grant-less transmissions. By developing an analytical framework based on a Markov chain representation of the LBT procedure, it shows that the GUL scheme is able to increase the uplink channel access probability in a MF system compared to a scheduled based scheme. In addition, along the paper system designs and details on how to enable this transmission scheme within the LTE ecosystem are elaborated.  Finally, comprehensive system level simulations are provided, and evaluation indicates that the proposed GUL mode can lead to a significant improvement of the uplink UPT performance with the negligible performance loss for MF downlink and Wi-Fi systems.

\balance

\end{document}